\documentclass[%
 reprint,
twocolomn,
 amsmath,amssymb,
 aps,
pra,
]{revtex4-2}

\usepackage{graphicx}
\usepackage{dcolumn}
\usepackage{bm}
\usepackage{bm}
\usepackage{color}

\begin{document}


\title{Many-hypercube codes: High-rate quantum error-correcting codes for high-performance fault-tolerant quantum computing}

\author{Hayato Goto}
 \email{hayato.goto@riken.jp}

\affiliation{%
 RIKEN Center for Quantum Computing (RQC), Wako, Saitama 351-0198, Japan
}%

\date{\today}

\begin{abstract}
Standard approaches to quantum error correction for fault-tolerant quantum computing are based on encoding a single logical qubit into many physical ones, resulting in asymptotically zero encoding rates and therefore huge resource overheads. To overcome this issue, high-rate quantum codes, such as quantum low-density parity-check codes, have been studied over the past decade. In this case, however, it is difficult to perform logical gates in parallel while maintaining low overheads. Here we propose concatenated high-rate small-size quantum error-detecting codes as a new family of high-rate quantum codes. Their simple structure allows for a geometrical interpretation using hypercubes corresponding to logical qubits. We thus call them \textit{many-hypercube codes}. They can realize both high rates, e.g., 30\% (64 logical qubits are encoded into 216 physical ones), and parallelizability of logical gates. Developing a high-performance decoder and encoders dedicated to the proposed codes, we achieve high error thresholds even in a circuit-level noise model. Thus, the many-hypercube codes will pave the way to high-performance fault-tolerant quantum computing.
\end{abstract}

\maketitle


\section{Introduction}

Quantum computers have been expected to outperform current classical computers by harnessing quantum superposition states. However, the quantum superpositions are notoriously fragile. The so-called decoherence leads to many errors in quantum computers, spoiling quantum computation. A standard approach to this issue is quantum error correction. Careful use of quantum error-correcting codes can protect quantum computation from errors, which is called fault-tolerant quantum computation~\cite{1,2,3}. A standard quantum error-correcting code for this purpose is the surface code~\cite{4,5,6,7,8,9,10}, which requires only a two-dimensional qubit array with nearest-neighbor interactions, and therefore is suitable for, e.g., superconducting-circuit implementations~\cite{11,12,13}. However, the surface code uses many physical qubits to protect a single logical qubit. More precisely, the number of physical qubits to encode a single logical qubit is the square of the code distance, where the code distance characterizes the code size and we can, in principle, correct arbitrary independent physical-qubit errors if the number of them is less than half the code distance. This means that the encoding rate defined as the ratio of the number of logical qubits to that of physical qubits vanishes rapidly as the code size becomes larger, resulting in large resource overheads~\cite{14,15,16}. 

In contrast, quantum low-density parity-check (qLDPC) codes~\cite{17} are known for their ability to achieve constant encoding rates, and therefore have been studied over the past decade to mitigate the above overhead issue~\cite{18,19}. Various kinds of qLDPC codes have been proposed~\cite{20,21,22,23,24,25,26,27,28}, high-performance decoders for them have also been developed~\cite{29,30,31,32,33,34,35}, and their physical implementations have recently been proposed~\cite{36,37,38,39,40}. However, they have relatively complex structure, making it difficult to implement logical gate operations in a fault-tolerant manner. A few methods for this purpose have been proposed~\cite{18,39,40,41,42,43}, but parallel execution of logical gates maintaining the advantage of the qLDPC codes, i.e., low overheads is still challenging.

Recently, a completely different approach to the constant encoding rate has been proposed~\cite{44}, which is based on a more conventional approach called code concatenation. Concatenation means recursive encoding with multiple codes. By increasing the encoding rates for higher concatenation levels, this proposal achieved a finite rate for the infinite code size. Importantly, this also allows for parallel execution of logical gates with constant overheads, unlike qLDPC codes. In other words, this proposal offers time-efficient, constant-space-overhead fault-tolerant quantum computation. However, this approach based on quantum Hamming codes has two issues. First, the encoding rate is not very high, converging to 1/36. Second, the decoding of this concatenated codes is based on hard-decision decoding, which is known to be suboptimal and has relatively low performance.

In this work, we propose another family of high-rate concatenated quantum codes. The characteristic feature of our proposal is the use of quantum error-detecting codes~\cite{2}, which have distance 2 and therefore can detect an error but cannot correct it. By concatenating the error-detecting codes, we can increase the code distance and thus obtain error-correcting codes. The advantage of the quantum error-detecting codes is their simple structure. Harnessing this advantage, Knill proposed concatenated quantum error-detecting codes called the ${C_4/C_6}$ scheme and achieved very high performance, i.e., the first realization of the error threshold exceeding 1\% in a circuit-level noise model~\cite{45}. (Recently, the concatenation of the ${C_4/C_6}$ scheme and the concatenated quantum Hamming codes mentioned above has been proposed to improve the performance of the latter~\cite{46}, but this still has the above-mentioned two issues, that is, the rate becomes rather lower and the decoding is still based on hard-decision decoding.) However, the ${C_4/C_6}$ scheme is based on a single-logical-qubit encoding, like the surface code, and therefore its encoding rate approaches zero rapidly as the code size increases. Unlike the ${C_4/C_6}$ scheme, the proposed concatenated codes have high encoding rates. In this work, we focus on the ${[\![6,4,2]\!]}$ code, which encodes four qubits into six qubits and has distance 2. (The reason for choosing this code is its relatively high rate and small size. The use of other codes, such as ${[\![4,2,2]\!]}$ and ${[\![8,6,2]\!]}$, is also interesting, but left for future work.) By concatenating it ${(L-1)}$ times, we obtain the ${[\![ N^{(L)},K^{(L)},D^{(L)}]\!]}={[\![6^L,4^L,2^L]\!]}$ code, which we refer to as the ${[\![6^L,4^L,2^L]\!]}$ level-$L$ many-hypercube code or simply the level-$L$ many-hypercube code for the reason explained later. Although it is not a constant-rate code, that is, its rate, $K^{(L)}/N^{(L)}={(4/6)^L}$, approaches zero as $L$ becomes larger, the rate is remarkably high for small $L$, e.g., 30\% (20\%) at the level 3 (4) with distance 8 (16), which is higher than not only the surface code, but also well-studied qLDPC codes~\cite{36,37,38,39,40}. Thus the many-hypercube codes will be promising as a near-term target. Note that the number of logical qubits is not limited to $K^{(L)}=4^L$. Using $M$ code blocks encoded with the level-$L$ many-hypercube codes as fault-tolerant quantum registers~\cite{44}, we can use $MK^{(L)}$ logical qubits for fault-tolerant quantum computing.

We developed a high-performance decoder dedicated to the many-hypercube codes based on level-by-level minimum distance decoding. (Our proposed decoding method will also be useful for other concatenated codes, such as the above concatenated quantum Hamming codes.) Using this decoder, we achieved a threshold of 5.6\% for bit-flip errors, which is comparable to the surface code (10.9\%)~\cite{47} and a 4\%-rate qLDPC (hypergraph product) code (7.5\%)~\cite{33}. We also propose fault-tolerant zero-state encoders for the many-hypercube codes. Using them, we achieved a threshold of 0.9\% for a logical controlled-NOT (CNOT) gate in a circuit-level noise model. Finally, we explain how to perform logical gates for the many-hypercube codes in parallel.

\section{Many-hypercube codes}

\begin{figure}[t]
    \includegraphics[width=\columnwidth]{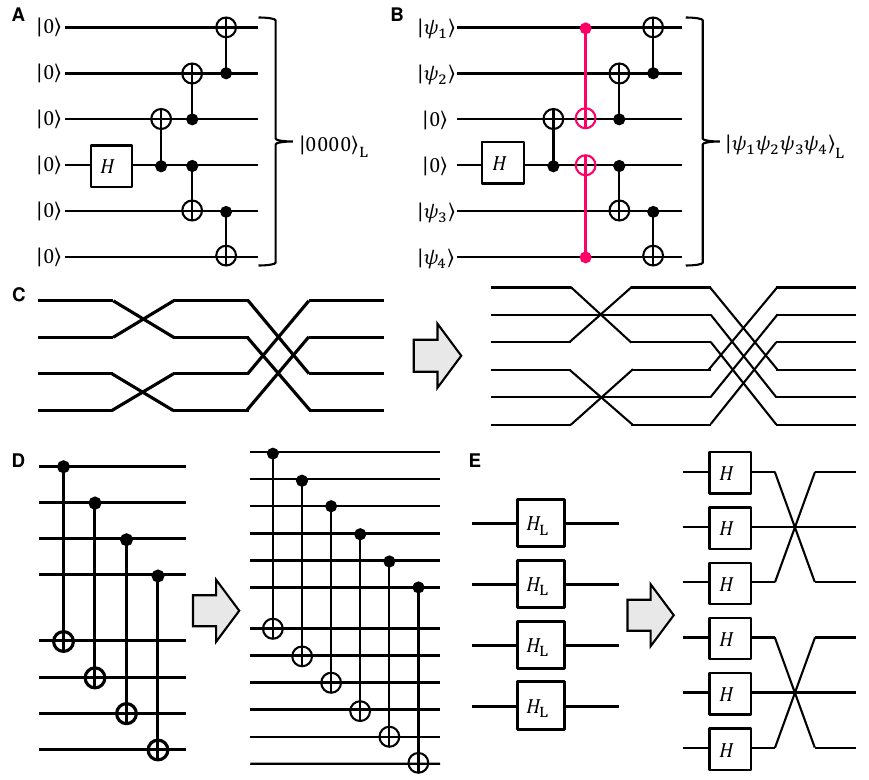}
    \caption{Encoders and logical gates of the ${[\![6,4,2]\!]}$ code. (A) Zero-state encoder. (B) Arbitrary-state encoder. In (B), additional two CNOT gates are highlighted in red. (C) Logical SWAP gates. (D) Transversal logical CNOT gates. (E) Transversal logical Hadamard gates. In (C--E), the left and right sides correspond to logical and physical ones, respectively.}
    \label{fig1}
\end{figure}

We start with the definition of the ${[\![ 6,4,2]\!]}$ code. 
The ${[\![ 6,4,2]\!]}$ code is one of the simplest stabilizer codes~\cite{3} and has only two stabilizer generators (check operators): ${S_Z=Z_1 Z_2 Z_3 Z_4 Z_5 Z_6}$ and ${S_X=X_1 X_2 X_3 X_4 X_5 X_6}$, which can detect an $X$ (bit-flip) error and a $Z$ (phase-flip) error, respectively (see Appendix~\ref{gate} for the definitions of elementary gates). Its four logical qubits are defined by the following logical $Z$ and $X$ operators:
\begin{align}
Z_{L1}&=Z_1 Z_2,~ X_{L1}=X_2 X_3,\\
Z_{L2}&=Z_2 Z_3,~ X_{L2}=X_1 X_2,\\
Z_{L3}&=Z_4 Z_5,~ X_{L3}=X_5 X_6,\\
Z_{L4}&=Z_5 Z_6,~ X_{L4}=X_4 X_5.
\end{align}
Other definitions are possible~\cite{2}, but we use this because of the geometrical interpretation of the code structure using hypercubes explained below. 
By the definition, the logical all-zero state of the ${[\![ 6,4,2]\!]}$ code is the six-qubit Greenberger-Horne-Zeilinger (GHZ) state: 
\begin{align}
|0000\rangle_L = \frac{|000000\rangle + |111111\rangle}{\sqrt{2}}.
\label{GHZ}
\end{align}
Therefore, the zero-state encoder of the ${[\![6,4,2]\!]}$ code is given by, e.g., the quantum circuit shown in Fig.~\ref{fig1}A. This can be generalized to an arbitrary-state encoder by adding two CNOT gates, as shown in Fig.~\ref{fig1}B.
The logical SWAP gates between the logical qubits 1 and 2, 3 and 4, and (1,2) and (3,4) can also be performed by physical SWAP gates between the physical qubits 1 and 3, 4 and 6, and (1,2,3) and (4,5,6), respectively, as shown in Fig.~\ref{fig1}C. 
We can also easily find that transversal logical CNOT and Hadamard gates can be performed by transversal physical CNOT and Hadamard (and SWAP) gates, respectively, as shown in Figs.~\ref{fig1}D and \ref{fig1}E.

\begin{figure}[t]
    \includegraphics[width=\columnwidth]{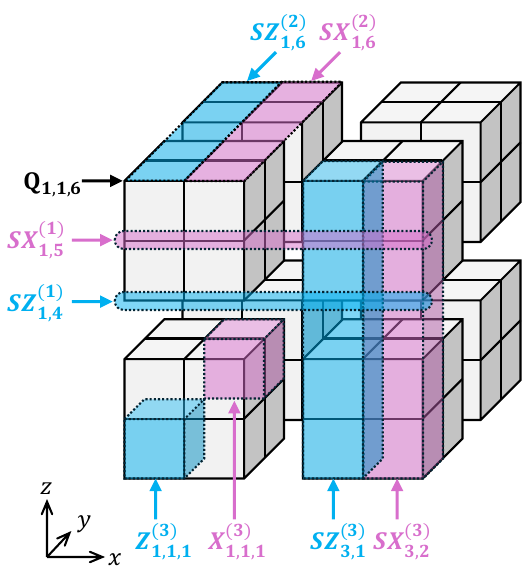}
    \caption{Visualization of the structure of the level-3 many-hypercube code. The vertices correspond to physical qubits. Examples of logical $Z$ and $X$ operators and stabilizers in Eqs.~(12)--(19) are highlighted by blue ($Z$) and red ($X$). In general, logical $Z$ and $X$ operators for the level-$L$ many-hypercube code correspond to $L$-dimensional hypercubes, the number of which is equal to that of the logical qubits. Also, level-$l$ stabilizers correspond to 
    $l$-dimensional objects on the hypercubes.}
    \label{fig2}
\end{figure}

Using the ${[\![6,4,2]\!]}$ code recursively, e.g., the level-3 many-hypercube code is defined as follows. Each of physical and encoded qubits is labelled by three integers (corresponding to the three levels) and denoted by Q with subscripts of the integers. The corresponding operators are also labelled similarly. The four level-1 qubits $\{ \mathrm{Q}_{i',j,k}^{(1)} | i'=1,\ldots,4 \}$ are encoded into the corresponding six physical qubits $\{ \mathrm{Q}_{i,j,k} | i=1,\ldots,6 \}$ with the ${[\![6,4,2]\!]}$ code ($j,k=1,\ldots,6$), where the two stabilizers and encoded $Z$ and $X$ operators are defined as follows: 
\begin{align}
SZ_{j,k}^{(1)}&=\bigotimes_{i=1}^6 Z_{i,j,k}, \\
SX_{j,k}^{(1)}&=\bigotimes_{i=1}^6 X_{i,j,k}, \\
Z_{1,j,k}^{(1)}&=Z_{1,j,k} Z_{2,j,k},~ X_{1,j,k}^{(1)}=X_{2,j,k} X_{3,j,k}, \\
Z_{2,j,k}^{(1)}&=Z_{2,j,k} Z_{3,j,k},~ X_{2,j,k}^{(1)}=X_{1,j,k} X_{2,j,k}, \\
Z_{3,j,k}^{(1)}&=Z_{4,j,k} Z_{5,j,k},~ X_{3,j,k}^{(1)}=X_{5,j,k} X_{6,j,k}, \\
Z_{4,j,k}^{(1)}&=Z_{5,j,k} Z_{6,j,k},~ X_{4,j,k}^{(1)}=X_{4,j,k} X_{5,j,k}.
\end{align}
Similarly, the four level-2 qubits $\{ \mathrm{Q}_{i',j',k}^{(2)} | j'=1,\ldots,4 \}$ are encoded into the corresponding six level-1 qubits $\{ \mathrm{Q}_{i',j,k}^{(1)} | j=1,\ldots,6 \}$ ($i'=1,\ldots,4, k=1,\ldots,6$) and the four level-3 qubits $\{ \mathrm{Q}_{i',j',k'}^{(3)} | k'=1,\ldots,4 \}$ are encoded into the corresponding six level-2 qubits $\{ \mathrm{Q}_{i',j',k}^{(2)} | k=1,\ldots,6 \}$ ($i',j'=1,\ldots,4$). In this case, we use the 64 level-3 encoded qubits as logical qubits in a fault-tolerant quantum register for quantum computing. In general, the level-$L$ many-hypercube code is defined similarly.

Interestingly, the above code structure can be visualized as shown in Fig.~\ref{fig2} by placing the physical qubit $\mathrm{Q}_{i,j,k}$ at the point with the coordinates ${(x,y,z)=(i,j,k)}$ in three-dimensional space, where the logical $Z$ and $X$ operators correspond to cubes. For example,
\begin{align}
Z_{1,1,1}^{(3)}
&=
Z_{1,1,1}^{(2)} Z_{1,1,2}^{(2)}
\nonumber \\
&=
Z_{1,1,1}^{(1)} Z_{1,2,1}^{(1)} 
Z_{1,1,2}^{(1)} Z_{1,2,2}^{(1)}
\nonumber \\
&=
Z_{1,1,1} Z_{2,1,1} Z_{1,2,1} Z_{2,2,1} 
Z_{1,1,2} Z_{2,1,2} Z_{1,2,2} Z_{2,2,2}
\\
X_{1,1,1}^{(3)}
&=
X_{1,1,2}^{(2)} X_{1,1,3}^{(2)}
\nonumber \\
&=
X_{1,2,2}^{(1)} X_{1,3,2}^{(1)} 
X_{1,2,3}^{(1)} X_{1,3,3}^{(1)}
\nonumber \\
&=
X_{2,2,2} X_{3,2,2} X_{2,3,2} X_{3,3,2} 
X_{2,2,3} X_{3,2,3} X_{2,3,3} X_{3,3,3}
\end{align}
are highlighted in blue and red, respectively, in Fig.~\ref{fig2}. Also, the level-$l$ stabilizers correspond to $l$-dimensional objects on the cubes, e.g., as follows: 
\begin{align}
SZ_{1,4}^{(1)}&=\bigotimes_{i=1}^6 Z_{i,1,4}, \\
SX_{1,5}^{(1)}&=\bigotimes_{i=1}^6 X_{i,1,5}, \\
SZ_{1,6}^{(2)}&=\bigotimes_{j=1}^6 Z_{1,j,6}^{(1)}
=\bigotimes_{j=1}^6 Z_{1,j,6} Z_{2,j,6}, \\
SX_{1,6}^{(2)}&=\bigotimes_{j=1}^6 X_{1,j,6}^{(1)}
=\bigotimes_{j=1}^6 X_{2,j,6} X_{3,j,6}, \\
SZ_{3,1}^{(3)}&=\bigotimes_{k=1}^6 Z_{3,1,k}^{(2)}
=\bigotimes_{k=1}^6 Z_{3,1,k}^{(1)} Z_{3,2,k}^{(1)}
\nonumber \\
&=\bigotimes_{k=1}^6 Z_{4,1,k} Z_{5,1,k} Z_{4,2,k} Z_{5,2,k}, \\
SX_{3,2}^{(3)}&=\bigotimes_{k=1}^6 X_{3,2,k}^{(2)}
=\bigotimes_{k=1}^6 X_{3,1,k}^{(1)} X_{3,2,k}^{(1)}
\nonumber \\
&=\bigotimes_{k=1}^6 X_{5,1,k} X_{6,1,k} X_{5,2,k} X_{6,2,k}.
\end{align}
which are also highlighted in blue ($SZ$) and red ($SX$) in Fig.~\ref{fig2}.

 In general, the logical $Z$ and $X$ operators of the level-$L$ many-hypercube code correspond to $L$-dimensional hypercubes in $L$-dimensional space and also its level-$l$ stabilizers correspond to $l$-dimensional objects on the hypercubes, where the number of the hypercubes is equal to that of the logical qubits. This is the reason for the name of the proposed code.
[The ``many" is used to distinguish it from a conventional hypercube code~\cite{48}, which is the ${[\![2^D,D,2]\!]}$ code defined on a single $D$-dimensional hypercube.]

The number of the vertices of each $L$-dimensional hypercube is $2^L$, which is equal to the code distance. That is, the size of the many-hypercube codes is increased by increasing not the edge length but the dimension while keeping the edge length. In contrast, the logical operators and stabilizers of topological codes, such as the surface code~\cite{4,5,6,7,8,9,10} and the color code~\cite{49,49,50,52}, correspond to the edges and faces, respectively, and their code sizes are increased by increasing the edge length.
Unlike the topological codes, the many-hypercube codes require interactions beyond nearest-neighbor ones. This is experimentally challenging, but recent experimental advances in ion-trap~\cite{53,53,54,55,56,57,58,59} and neutral-atom~\cite{60,61,62,63} systems are encouraging. In fact, hypercube connectivity has already been realized experimentally~\cite{63}.

\section{Decoders}

In this work, we investigated three decoding methods for the many-hypercube codes. The first one is Knill’s method proposed for the ${C_4/C_6}$ scheme~\cite{45}, which we refer to as hard-decision decoding. The second one is soft-decision decoding based on a posteriori probability calculation~\cite{64,65}. The original soft-decision decoding is based on block-MAP (maximum a posteriori probability) decoding, which cannot be applied directly to high-rate codes such as the many-hypercube codes, because then we must calculate a posteriori probabilities for an exponentially large number of codewords. We thus modified it to symbol-MAP decoding applicable to high-rate codes. The third one is our proposed method based on level-by-level minimum distance decoding, where we keep only minimum-distance codewords from measurement outcomes as candidates at each level from level 1 to the logical level. See Appendices~\ref{hard}--\ref{MD} for the details of the three methods.

\begin{figure}[h]
    \includegraphics[width=\columnwidth]{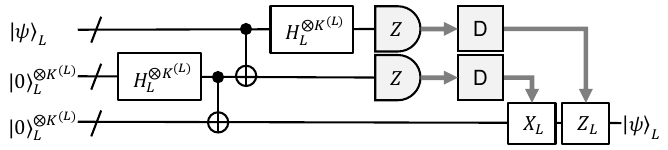}
    \caption{Error-correcting teleportation. The input logical-qubit state encoded with the level-$L$ many-hypercube code, ${|\psi \rangle_L}$, on the left side is teleported with two logical all-zero states to the right side. The gray bold arrows show the flow of the classical information. The “D” boxes are the decoders which decode the physical-qubit measurement outcomes and provide logical-qubit ones. If a logical-qubit outcome is 1, the logical $Z$ or $X$ operator is performed on the corresponding logical qubit. The transversal logical CNOT and Hadamard gates are performed by transversal physical CNOT and Hadamard gates, as shown in Figs.~\ref{fig1}D and \ref{fig1}E, respectively.}
    \label{fig3}
\end{figure}

In the following, we assume that error correction is done by error-correcting teleportation (ECT)~\cite{45,65,66}. 
This is quantum teleportation with logical qubits, where the classical feed-forward information is determined by decoding physical-qubit measurement outcomes, as shown in Fig.~\ref{fig3}. Thus, the decoding is the task to estimate the measurement outcomes of logical qubits from those of physical qubits in the logical Bell measurements. Note that in ECT, we obtain all the syndrome information at once, no need to repeatedly measure them one by one, unlike the Shor method~\cite{1,18} used for the surface code and qLDPC codes. Assuming that the two ancilla registers used for the quantum teleportation are reliable sufficiently, the errors in ECT come mainly from the decoding error. Also note that all the errors to be decoded are included in the physical-qubit measurement errors in the Bell measurement, and therefore in the case of ECT, decoders designed for independent physical-qubit errors are directly applicable to circuit-level noise models. In this work, we also assume that all the measurements are done in the $Z$ basis ${\{ |0\rangle, |1\rangle \}}$, resulting in bit-value outcomes.

We numerically evaluated the performance of the three decoding methods using a bit-flip error model (see Appendix~\ref{bit-flip simulation} for details). In this work, stabilizer quantum-circuit simulations in our numerical study were done using a python package called Stim~\cite{67}. The decoding error probabilities estimated by the simulations are shown in Fig.~\ref{fig4}. First, the exponents of the power-function fits for the hard-decision decoding are smaller than half the code distance, as in the case of the ${C_4/C_6}$ scheme~\cite{65}, which shows its suboptimality. On the other hand, the exponents for the other two are near to or even exceed half the distance, showing their high performance. Second, the error thresholds are estimated at 1.1\%, 1.5\%, and 5.6\%, respectively, and only our minimum distance decoding is comparable to the surface code (10.9\%)~\cite{47} and a well-studied qLDPC (4\%-rate hypergraph product) code (7.5\%)~\cite{33}. Thus our proposed decoding method achieves the highest performance among the three methods.

\section{Fault-tolerant zero-state encoders}

\begin{widetext}
    
    \begin{figure}[t]
        \includegraphics[width=\columnwidth]{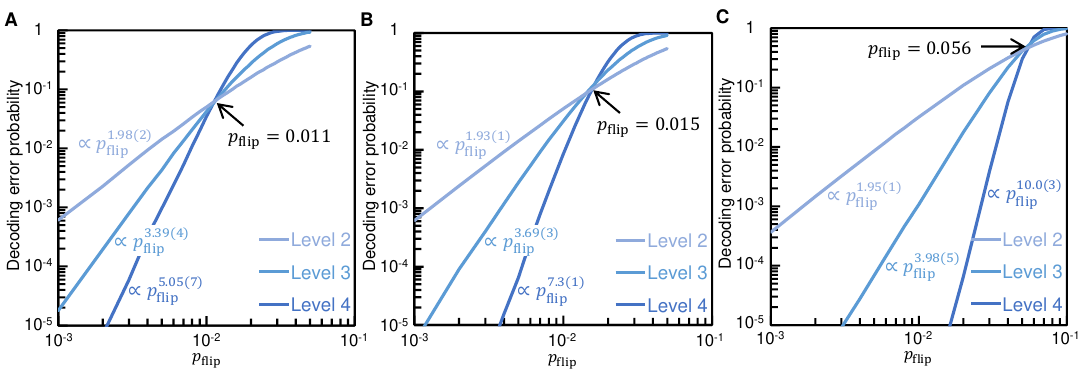}
        \caption{Performance of three decoders of the many-hypercube codes for bit-flip errors. (A) Hard-decision decoder. (B) Symbol-MAP decoder. (C) Level-by-level minimum distance decoder. The exponents are estimated by fitting a power function to the linear parts of the log-log plots. The thresholds indicated by arrows are defined by the intersection points of the level-3 and level-4 curves.
        See Appendices~\ref{hard}--\ref{MD} for the details of the three decoding methods and Appendix~\ref{bit-flip simulation} for the simulation.}
        \label{fig4}
    \end{figure}

    \begin{figure}[t]
        \includegraphics[width=0.8\columnwidth]{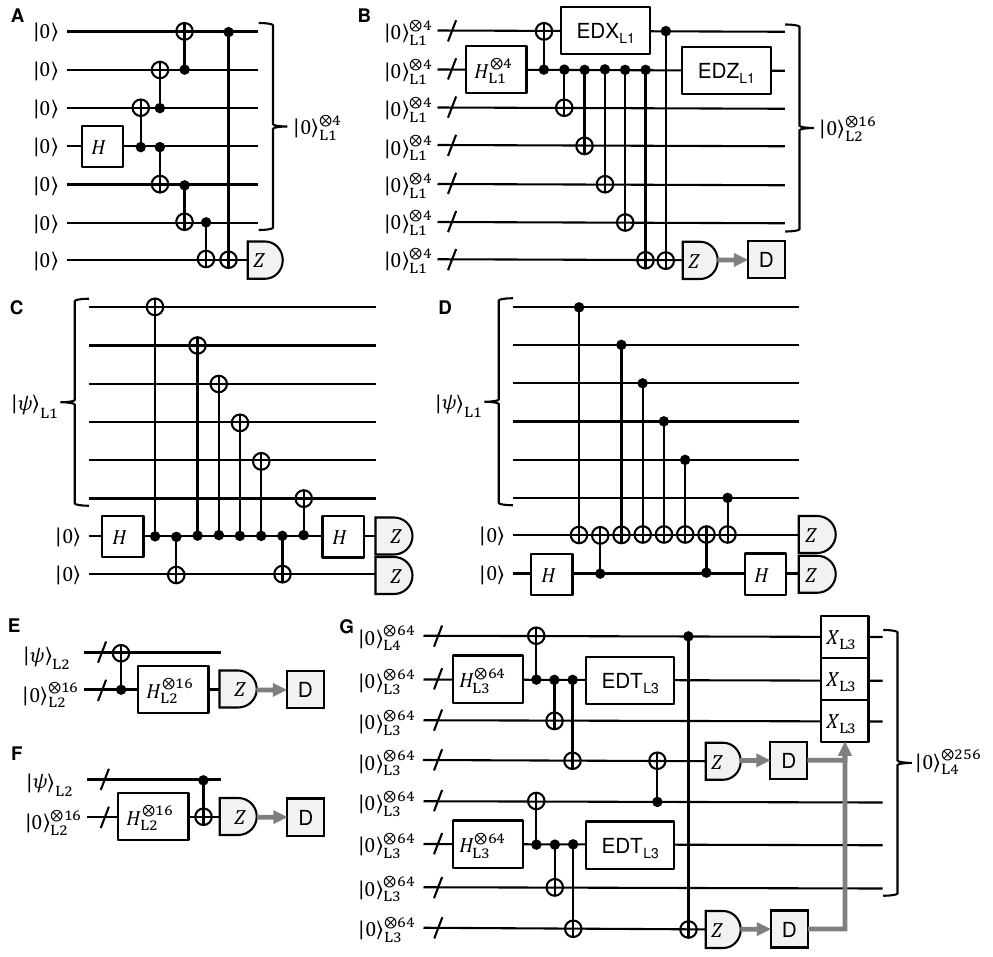}
        \caption{Fault-tolerant zero-state encoders of the many-hypercube codes. (A) Level-1 zero-state encoder. The inputs are seven physical qubits in zero states. The CNOT and Hadamard gates are physical ones. The encoding is repeated until the measurement outcome of the ancilla qubit is 0. (B) Level-2 zero-state encoder. The transversal level-1 encoded CNOT and Hadamard gates are performed by transversal physical gates, as in Figs.~\ref{fig1}D and \ref{fig1}E. The measurement outcomes of the level-1 encoded ancilla qubits are obtained by decoding the physical-qubit outcomes with the hard-decision decoding for error detection (see Appendix~\ref{hard}). The encoding is repeated until no error is detected and all the level-1 outcomes are 0. (C and D) Level-1 $Z$-error and $X$-error detection gadgets (EDZ$_{L1}$ and EDX$_{L1}$), respectively, used for the level-2 zero-state encoder in (B). Errors are detected unless both the measurement outcomes of the two ancilla qubits are 0. (E and F) Level-2 $Z$-error and $X$-error detection gadgets (EDZ$_{L2}$ and EDX$_{L2}$), respectively, used for the level-3 zero-state encoder. Errors are detected unless none of the outcomes of the minimum distance decoding for error detection with $L_D=1$ is $F$ (see Appendix~\ref{MD}). Level-3 zero-state encoder is similar to (B) except replacing EDZ$_{L1}$ and EDX$_{L1}$ with EDZ$_{L2}$ and EDX$_{L2}$. (G) Level-4 zero-state encoder. EDT$_{L3}$ denotes a level-3 error-detecting teleportation gadget, where the minimum distance decoding for error detection with $L_D=2$ is used (see Appendix~\ref{MD}). The same decoding is used for decoding the measurement outcomes of the two ancilla registers. The encoding is repeated until no error is detected and all the parities of the outcomes of the level-3 encoded-qubit pairs in the two ancilla registers are even, where if both the outcomes of a pair are 1, we perform encoded $X$ gates on either of the first or second half of the six level-3 encoded qubits.}
        \label{fig5}
    \end{figure}
\end{widetext}

Hereafter, we consider a circuit-level noise model with error rate $p_{\mathrm{circ}}$. In this work, we assume the following noise model. Each of physical-qubit zero-state preparation and $Z$-basis measurement is accompanied by a bit flip with probability $p_{\mathrm{circ}}$, and each two-qubit (CNOT) gate is followed by one of 15 two-qubit Pauli errors with equal probability $p_{\mathrm{circ}}/15$~\cite{8}. 
On the other hand, we assume no single-qubit-gate and memory errors for simplicity. This model is relevant for ion-trap~\cite{53,54,55,56,57,58,59} and neutral-atom~\cite{60,61,62,63} systems. 
(Even for these systems, it is eventually desirable to consider the effect of the neglected errors, but it requires thorough optimization of time overheads and therefore is left for future work.)

In such a circuit-level noise model, we cannot use the non-fault-tolerant encoders in Figs.~\ref{fig1}A and \ref{fig1}B, because of undetectable correlated errors due to CNOT gates in a code block. Therefore we modify the zero-state encoder at each level as follows. In the following, we use various kinds of error detection and repeat encoding until no error is detected.

For the encoding of the level-1 all-zero state, which is the six-qubit GHZ state as shown in Eq.~(\ref{GHZ}), we can use the well-known fault-tolerant GHZ-state preparation method with an ancilla qubit shown in Fig.~\ref{fig5}A. We accept the encoding if the measurement outcome of the ancilla qubit is 0, otherwise reject and restart it from the beginning. Thus logical $X$ errors can be eliminated. (Logical $Z$ errors do not occur on the logical all-zero state.) To evaluate the space and time overheads for the zero-state encoders, here we introduce the total number of physical qubits including ancilla ones, $N^{\prime (L)}$, and the circuit depth including state preparation and measurement, $T^{(L)}$, respectively, required for the level-$L$ zero-state encoder. From Fig.~\ref{fig5}A, 
we obtain
$N^{\prime (1)}=7$ and $T^{(1)}=8$.

We cannot apply this method directly to level 2, because of uncorrectable intra-block errors. Also, the low-depth circuit for the GHZ-state preparation in Fig~\ref{fig5}A results in many error-detection gadgets to eliminate the intra-block $Z$ errors. To achieve fault tolerance with minimum effort at level 2, we propose the level-2 zero-state encoder shown in Fig.~\ref{fig5}B, where the control qubits of the CNOT gates concentrate on a single qubit in order to eliminate the intra-block $Z$ errors by a single error-detection gadget. To eliminate intra-block and logical $X$ errors, we also need only a single error-detection gadget and ancilla-qubit measurement. To minimize space overheads, the level-1 error-detection gadgets are implemented by the flag-based method with two physical ancilla qubits and depth 12 in Figs.~\ref{fig5}C and \ref{fig5}D~\cite{68}. The transversal encoded CNOT and Hadamard gates can be performed fault-tolerantly by transversal physical gates with depth 1 and 2, respectively, as shown in Figs.~\ref{fig1}D and \ref{fig1}E. Optimizing the overlaps of physical operations to minimize the time overhead, we obtain
$N^{\prime (2)}=N^{\prime (1)}\times 7 + 2\times 2=53$ and $T^{(2)}=T^{(1)}+2+6+(12-2)=26$.
Note that the space overhead, i.e., the total number of physical qubits, can be reduced by reusing the ancilla qubits, but then the time overhead will increase. Also note that we can use the Steane method with an encoded ancilla qubit~\cite{69} for the error-detection gadgets, which in the level-2 case are shown in Figs.~\ref{fig5}E and \ref{fig5}F, but this results in larger space overheads and no performance improvement (see Appendix~\ref{sec-level2}). This is because at level 1, we only need to measure just a single weight-6 stabilizer, which can be achieved most easily by the flag-based method.

The level-3 zero-state encoder is the same as the level-2 one in Fig.~\ref{fig5}B if the levels are raised by one, but level-2 error-detection gadgets are implemented by the above-mentioned Steane method~\cite{69} shown in Figs.~\ref{fig5}E and \ref{fig5}F. This is because at level 2, we need to measure six weight-6 level-1 stabilizers and four weight-12 level-2 stabilizers, which can be achieved by the Steane method more easily than the Shor method with physical ancilla qubits. The space and time overheads at level 3 are estimated as 
$N^{\prime (3)}=N^{\prime (2)}\times 7 + N^{\prime (2)}\times 2=477$ and $T^{(3)}=T^{(2)}+2+6+4=38$.
where and in the following, decoding is not counted for time-overhead evaluation.

Although the encoder in Fig.~\ref{fig5}B can also be applied to the level-4 case, then the acceptance probabilities at error-detection gadgets become too low if we use the most stringent condition for error detection. To mitigate this issue, we can relax the error-detection condition (see Appendix~\ref{MD}), but then the logical gate performance becomes low (see Appendix~\ref{encoder5B} for detailed results using the encoder in Fig.~\ref{fig5}B at level 4). To improve the performance, we propose the level-4 zero-state encoder shown in Fig.~\ref{fig5}G, which uses two level-3 four-qubit GHZ states with error-detecting teleportation (EDT) gadgets eliminating intra-block $Z$ and $X$ errors simultaneously. The reason why this encoder can achieve higher performance than that in Fig.~\ref{fig5}B is that the intra-block errors in two four-qubit GHZ states are independent and therefore detectable with high probability by the ancilla measurements. From Fig.~\ref{fig5}G, the space and time overheads at level 4 are estimated as 
$N^{\prime (4)}=N^{\prime (3)}\times 8 + N^{\prime (3)}\times 2 \times 2=5724$ and $T^{(4)}=T^{(3)}+2+3+4=47$.
where and in the following, Pauli operations are not counted for time-overhead evaluation.

As numerically shown below, the encoders in Fig.~\ref{fig5} well satisfy fault tolerance. The time overheads evaluated above are almost proportional to the level. (The effect of postselection at error-detection gadgets is discussed later.) On the other hand, the logical-gate error rates decrease doubly exponentially with respect to the level, as numerically shown later. These facts suggest that the encoding only needs doubly logarithmic time overheads with respect to computational size and hence is time efficient~\cite{44}. 
Although the space overheads increase more rapidly, the net encoding rates defined by $K^{(L)}/N^{\prime (L)}$ including ancilla qubits are still relatively high, higher than 4\% even at level 4. 
(The encoding rate for the same-distance surface code is $1/16^2=0.4$\% even without ancilla qubits.) 
Further optimization of the space and time overheads may be possible but is left for future work.

The comparison with the $C_4/C_6$ scheme~\cite{45,65} is as follows. 
The parameters of the $C_4/C_6$ scheme are given by $N^{(L)}=4\times 3^{L-1}$, $K^{(L)}=1$ (or 2), $D^{(L)}=2^L$, and $N^{\prime (L)}=4 \times 12^{L-1}$~\cite{65}. 
By the present technique for the many-hypercube codes, the space overheads may be reduced to $N^{\prime (L)}=5^L$. Then the net encoding rate at level 4 is $K^{(4)}/N^{\prime (4)}=0.16$\% (or 0.32\%), 
which is much lower than that of the level-4 many-hypercube code, as expected. On the other hand, the logical CNOT performance of the $C_4/C_6$ scheme~\cite{65} is much higher than that of the many-hypercube codes presented below. This may be due to the high rates of the many-hypercube codes and the optimal (block-MAP) decoder for the $C_4/C_6$ scheme.

\vspace{1cm}

\section{Logical gate operations}

Arbitrary encoded Pauli gates can easily be performed fault-tolerantly by physical Pauli gates (or the so-called Pauli frame~\cite{45,66}) according to the definitions of the encoded Pauli operators, e.g., in Eqs.~(12) and (13).

Logical SWAP gates can also be performed easily by physical SWAP gates or renumbering of physical qubits as follows. The level-1 encoded SWAP gates are performed as shown in Fig.~\ref{fig1}C. 
At higher levels, e.g., at level 3, simultaneous physical SWAP gates 
$\{ Q_{1,j,k} \leftrightarrow Q_{3,j,k} |j,k=1,\ldots,6 \}$ result in simultaneous logical SWAP gates $\{ Q_{1,j',k'}^{(3)} \leftrightarrow Q_{2,j',k'}^{(3)} |j',k'=1,\ldots,4 \}$. Similarly, $\{ Q_{i',3,k'}^{(3)} \leftrightarrow Q_{i',4,k'}^{(3)} |i',k'=1,\ldots,4 \}$ and $\{ (Q_{i',j',1}^{(3)},Q_{i',j',2}^{(3)}) \leftrightarrow (Q_{i',j',3}^{(3)},Q_{i',j',4}^{(3)} )|i',j'=1,\ldots,4 \}$ are performed by corresponding simultaneous physical SWAP gates, $\{ Q_{i,4,k} \leftrightarrow Q_{i,6,k} |i,k=1,\ldots,6 \}$ and $\{ (Q_{i,j,1},Q_{i,j,2},Q_{i,j,3} ) \leftrightarrow (Q_{i,j,4},Q_{i,j,5},Q_{i,j,6} )|i,j=1,\ldots,6 \}$, respectively. In general, we can perform simultaneous logical SWAP gates between the logical qubits corresponding to the hypercubes on two parallel hyperplanes by simultaneous physical SWAP gates.

For the other gates necessary for universal computation (see Appendix~\ref{gate}), we first consider the case where the same gate is performed on all the encoded/physical qubits in a code block, which we refer to as full transversal gates. If the same gate is performed on only some specified qubits in a code block, we call it partial transversal gates, which is harder to implement.

\begin{figure}[t]
    \includegraphics[width=\columnwidth]{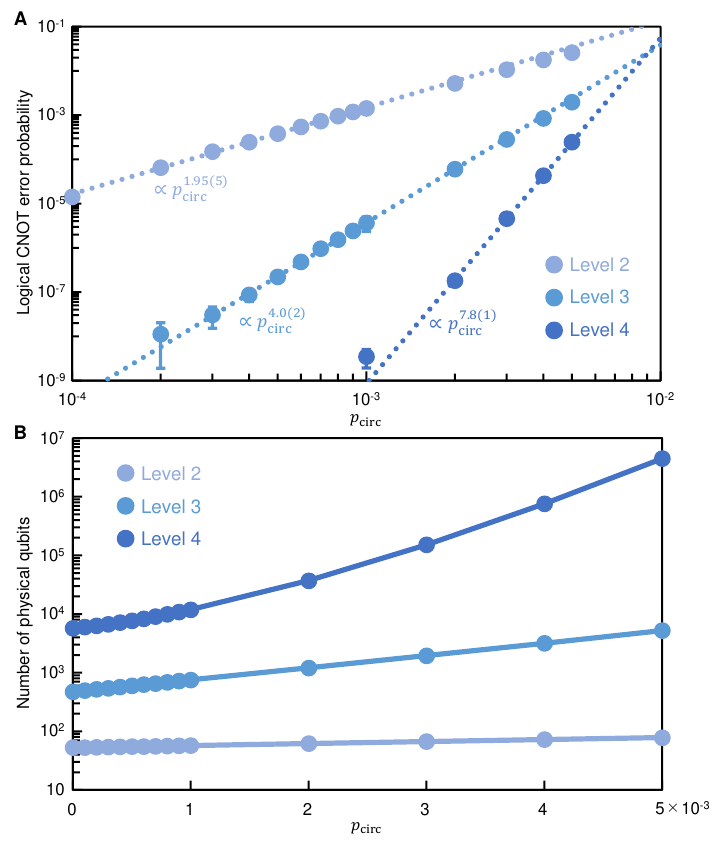}
    \caption{Performance of full transversal logical CNOT gates for the many-hypercube codes in the circuit-level error model. (A) Each circle shows the error probability per logical CNOT gate for error rate $p_{\mathrm{circ}}$. The dotted lines are power-function fits to the five data points from the lowest, from which the exponents are estimated. 
    (B) The number of physical qubits (space overhead) required for the preparation of a logical all-zero state.
    See Appendix~\ref{bit-flip simulation} for the details of the simulation and the main text for the definition of the error model.}
    \label{fig6}
\end{figure}

From Figs.~\ref{fig1}D and \ref{fig1}E, full transversal logical CNOT and Hadamard gates can be performed fault-tolerantly by full transversal physical CNOT and Hadamard gates (and physical SWAP gates) followed by ECT gadgets. Their time overheads are dominated by the ECT gadgets, which are estimated at $T^{(L)}+7$ in the level-L case from Fig.~\ref{fig3}, and therefore time efficient. We numerically evaluated the performance of the full transversal logical CNOT gates using the above-mentioned circuit-level noise model (see Appendix~\ref{CNOT simulation} for details). The logical CNOT error probabilities estimated by the simulation are shown in Fig.~\ref{fig6}A. The error threshold is estimated at 0.9\%, which is comparable to the surface code (1.1\%)~\cite{70} and a recently developed qLDPC quantum memory (0.7\%)~\cite{40}. Also, the exponents of the power-function fits are close to half the code distance, showing that the logical CNOT gates (and the zero-state encoders in Fig.~\ref{fig5}) well satisfy fault tolerance. (The level-4 value at $p_{\mathrm{circ}}=10^{-3}$ seems an outlier, resulting in a smaller exponent, e.g., 7, in the range of $p_{\mathrm{circ}}\le 10^{-3}$ than that estimated by fitting, 7.8, shown in Fig.~\ref{fig6}A. The exponent smaller than half the code distance may be due to the relaxation of the error-detection condition mentioned above.)

In the above simulation of logical CNOT gates, we also estimated the expectation value of the total number of physical qubits to prepare a logical all-zero state $|0\rangle_L^{\otimes K^{(L)}}$ taking the effect of postselection and restart into account. The results are shown in Fig.~\ref{fig6}B. It turns out that we need $p_{\mathrm{circ}}\le 10^{-3}$ to achieve the increasing rate less than 2 even at level 4. This will be achievable for ion-trap and neutral-atom systems~\cite{38}. [In Fig.~\ref{fig6}B, the numbers at $p_{\mathrm{circ}}=0$ are $N^{\prime (L)}$ in Eqs.~(20)--(23).]

Next, we consider full transversal logical phase and non-Clifford gates, $S$ and ${R_Y (\pi/4)}$ (see Appendix~\ref{gate} for their definitions). It is known that these two gates can be performed using ancilla qubits in $|Y\rangle=(|0\rangle + i|1\rangle)/\sqrt{2}$ and $|H\rangle=\cos (\pi/8) |0\rangle + \sin (\pi/8) |0\rangle$, respectively, together with CNOT and Hadamard gates (see Figs.~\ref{fig7}A and \ref{fig7}B), where $|Y\rangle$ and $|H\rangle$ are the eigenstates of $Y$ and $H$ with eigenvalue 1, respectively~\cite{15,71}. Therefore the full transversal logical $S$ and ${R_Y (\pi/4)}$ require to fault-tolerantly prepare the level-$L$ logical $|Y\rangle_L^{\otimes K^{(L)}}$ and $|H\rangle_L^{\otimes K^{(L)}}$, which can be achieved by non-fault-tolerant encoding with the arbitrary-state encoder in Fig.~\ref{fig1}B followed by state distillation and level-raising teleportation, as originally proposed for the $C_4/C_6$ scheme~\cite{71}. (The full transversal logical non-Clifford gates also need partial transversal logical Hadamard gates depending on the measurement outcomes, which are explained later.) We propose the 2-to-1 distillation method for $|Y\rangle_L^{\otimes K^{(L)}}$ based on $HS|Y\rangle=|1\rangle$ shown in Fig.~\ref{fig7}C. Note that our method needs only two noisy $|Y\rangle_L^{\otimes K^{(L)}}$, unlike the well-known 7-to-1 distillation method with the Steane code~\cite{15}. This difference comes from the fact that the many-hypercube codes are based on distance-2 codes, and therefore do not need distance-3 codes, such as the Steane code, for distillation. For $|H\rangle_L^{\otimes K^{(L)}}$, we can use the standard 14-to-2 distillation method based on the distance-2 $H_6$ code~\cite{71,72}. The level-raising teleportation shown in Fig.~\ref{fig7}D can convert four distilled $|Y\rangle_{L-1}^{\otimes K^{(L-1)}}$ and 
$|H\rangle_{L-1}^{\otimes K^{(L-1)}}$ to $|Y\rangle_L^{\otimes K^{(L)}}$ and $|H\rangle_L^{\otimes K^{(L)}}$, respectively, in a fault-tolerant manner~\cite{71}.

In the following, we consider partial transversal logical gates. 
The partial transversal logical Hadamard gates can be performed using an ancilla register where some logical qubits are in $|+\rangle=H|0\rangle=(|0\rangle+|1\rangle)/\sqrt{2}$ and the others in $|0\rangle$, which we refer to as a partial $|+\rangle$ state, together with full transversal logical CNOT and Hadamard gates and an ECT gadget, as shown in Fig.~\ref{fig8}A. 
(The ECT gadget is necessary to prevent intra-block errors from spreading through the logical CNOT and Hadamard gates.) Logical partial $|+\rangle$ states can be prepared fault-tolerantly in a similar manner to $|Y\rangle_L^{\otimes K^{(L)}}$ and $|H\rangle_L^{\otimes K^{(L)}}$ explained above. At level 1, we can fault-tolerantly prepare encoded partial $|+\rangle$ states more efficiently than state distillation, as follows. Any level-1 encoded partial $|+\rangle$ state can be prepared by one of the three methods in Figs. 8B--8D. The first and second encoders in Figs.~\ref{fig8}B and \ref{fig8}C are based on the fact that some level-1 partial $|+\rangle$ states can be expressed with only the Bell state, $(|00\rangle+|11\rangle)/\sqrt{2}$, and the four-qubit GHZ state by definition. The third encoder in Fig.~\ref{fig8}D is obtained from the arbitrary encoder in Fig.~\ref{fig1}B followed by verification with minimum effort through the measurements of an encoded $Z$ operator and an encoded $X$ operator, in a similar manner to the most efficient fault-tolerant Steane-code encoder with a single ancilla qubit~\cite{73}. Applying the level-raising teleportation to the resultant level-1 partial $|+\rangle$ states, we obtain level-2 partial $|+\rangle$ states. At higher levels, we use the 4-to-1 distillation method shown in Fig.~\ref{fig8}E.

The total number of physical qubits and circuit depth for the preparation of a level-$L$ logical partial $|+\rangle$ state, which are denoted by $N_+^{(L)}$ and $T_+^{(L)}$, respectively, are estimated from Figs.~\ref{fig7}D, \ref{fig8}D, and \ref{fig8}E in the worst-case scenario where all the level-1 blocks are encoded by the encoder in Fig.~\ref{fig8}D as follows:
$N_+^{(1)}=8$, $T_+^{(1)}=9$, 
$N_+^{(2)}=N_+^{(1)}\times 4 + N^{\prime (2)}\times 2=138$, $T_+^{(2)}=T_+^{(1)}+6+4=20$, 
$N_+^{(3)}=N_+^{(2)}\times 4\times 4+N^{\prime (3)} \times 2=3162$, $T_+^{(3)}=T_+^{(2)}+6+4=30$,
$N_+^{(4)}=N_+^{(3)}\times 4\times 4+ N^{\prime (4)} \times 2=62040$, and $T_+^{(4)}=T_+^{(3)}+6+4=40$.
Although this preparation is time efficient, the space overheads are considerably larger than those for the encoded all-zero states. More efficient encoding of partial $|+\rangle$ states may be possible but left for future work.

From Figs.~\ref{fig7}A and \ref{fig7}B, we can perform partial transversal logical $S$ and $R_Y (\pi/4)$ by replacing the full transversal logical Hadamard gates with partial ones explained above, because then the two logical CNOT gates on the qubits without logical Hadamard gates are cancelled out.

\begin{widetext}

\begin{figure}[t]
    \includegraphics[width=0.8\columnwidth]{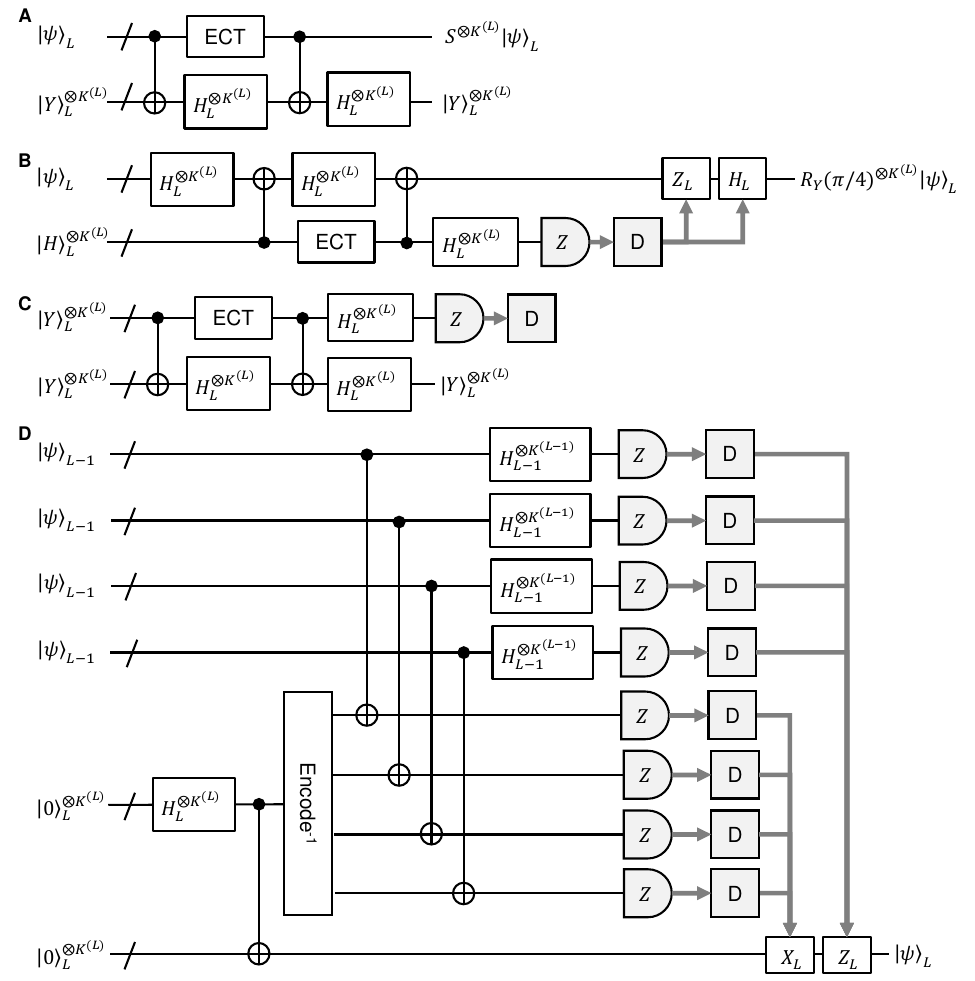}
    \caption{Logical phase and non-Clifford gates. (A and B) Full transversal logical phase and non-Clifford gates, $S$ and ${R_Y (\pi/4)}$, respectively. In (B), the measurement outcomes are decoded, and the logical $Z$ and $H$ are performed if the corresponding decoding outcome is 1. ECT is inserted between two CNOT gates for preventing intra-block errors from spreading due to physical SWAP gates in the logical Hadamard gate. Partial transversal logical phase and non-Clifford gates can be performed by replacing the full transversal logical Hadamard gates with partial ones in Fig.~\ref{fig8}A. 
    (C) Proposed 2-to-1 distillation for $|Y\rangle_L^{\otimes K^{(L)}}$. The distillation is accepted if all the decoding outcomes are 1. 
    (D) Level-raising teleportation. The box “Encode$^{-1}$” is the gadget performing the inversion of the arbitrary-state encoder in Fig. 1B, outputting four level-($L$-1) states, where the other two level-($L$-1) blocks are measured and decoded by an error-detection decoder, and the Bell-state preparation is repeated until all the decoding outcomes are 0. In the Bell measurements, the measurement outcomes are also decoded by an error-detection decoder~\cite{71}.}
    \label{fig7}
\end{figure}

\begin{figure}[t]
    \includegraphics[width=0.9\columnwidth]{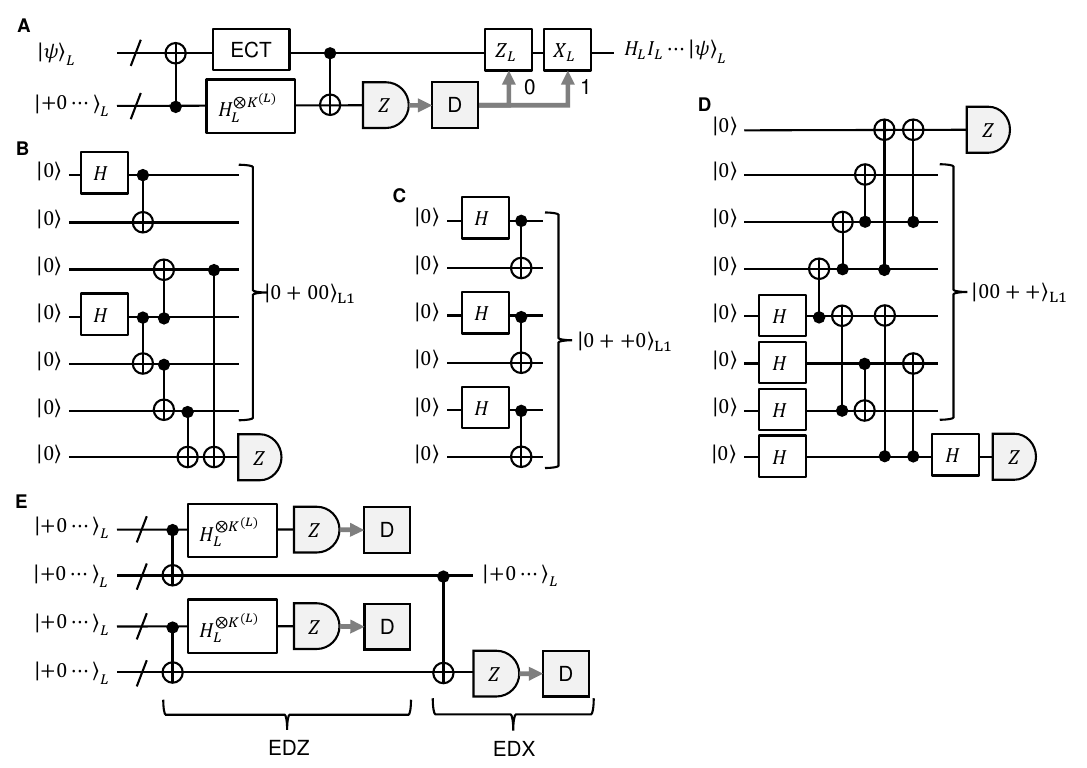}
    \caption{Partial transversal logical Hadamard gates. (A) Partial transversal logical Hadamard gates with a logical partial $|+\rangle$ state and full transversal logical CNOT and Hadamard gates. The logical $Z$ ($X$) is performed if the decoding outcome is 0 (1). ECT is inserted between two CNOT gates for preventing errors from spreading due to physical SWAP gates in the logical Hadamard gate. (B--D) Fault-tolerant preparation methods for level-1 logical partial $|+\rangle$ states. The other ones can also be prepared similarly to one of the three through SWAP gates and full transversal Hadamard gates. (E) 4-to-1 distillation for a logical partial $|+\rangle$ state. This is repeated until all the decoding outcomes are consistent with the input state.}
    \label{fig8}
\end{figure}

\begin{figure}[t]
    \includegraphics[width=0.75\columnwidth]{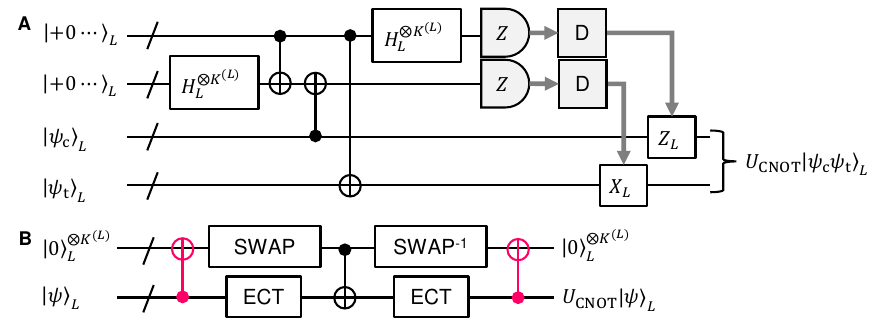}
    \caption{Logical CNOT gates. (A) Partial transversal logical CNOT gates (inter-block logical CNOT gates) with two ancilla registers in the same partial $|+\rangle$ states (showing only two logical qubits) and full transversal logical CNOT and Hadamard gates. The $Z$ and $X$ are performed if the corresponding decoding outcome is 1. Then, the logical CNOT gates denoted by $U_{\mathrm{CNOT}}$ are performed on the logical qubits corresponding to $|+\rangle$ of the partial $|+\rangle$ state in the control and target states, $|\psi_c \rangle_L$ and $|\psi_t \rangle_L$, respectively. (B) Intra-block logical CNOT gates using an ancilla register in the logical all-zero state. The first and third CNOT gates highlighted in red are partial transversal logical CNOT gates on the control qubits of the logical CNOT gates to be performed on $|\psi \rangle_L$. (The second one is full transversal logical CNOT gates.) The SWAP gate moves the logical qubits corresponding to the control qubits to the ones corresponding to the target qubits of the logical CNOT gates to be performed. They are returned to the original positions by the SWAP$^{-1}$ gate. Thus the desired intra-block logical CNOT gates denoted by $U_{\mathrm{CNOT}}$ are performed on $|\psi \rangle_L$.}
    \label{fig9}
\end{figure}

\end{widetext}

Partial transversal logical CNOT gates between two registers (inter-block logical CNOT gates) can be performed using two ancilla registers in the same partial $|+\rangle$ state and full transversal logical CNOT and Hadamard gates, as shown in Fig.~\ref{fig9}A. In this case, each logical CNOT gate is performed on the corresponding two logical qubits with the same label in the two registers. Using logical SWAP gates explained above, we can perform at least a logical CNOT gate on an arbitrary logical-qubit pair between the two registers. Finally, logical CNOT gates in a register (intra-block logical CNOT gates) can be performed with an ancilla register in the logical all-zero state, logical SWAP gates, and inter-block full and partial transversal logical CNOT gates, as shown in Fig.~\ref{fig9}B.

Thus, we can implement a universal gate set for the many-hypercube codes fault-tolerantly and perform most of them in parallel. For qLDPC codes, we can, in principle, achieve parallel execution of logical gates in a similar manner, but then logical ancilla states required for it are difficult to prepare, which may need another high-threshold code such as concatenated Steane codes~\cite{18}. On the other hand, for high-rate concatenated codes, such as the many-hypercube codes and the concatenated quantum Hamming codes~\cite{44}, the required ancilla states themselves are encoded with the same high-rate concatenated codes and therefore relatively easy to prepare. This is the advantage of high-rate concatenated codes over qLDPC codes. To overcome the ancilla-preparation issue, a current standard approach to logical gates for qLDPC codes uses the teleportation between a logical qubit in a qLDPC quantum memory and another logical qubit encoded with, e.g., the surface code~\cite{39,40}, where logical gates are performed on the surface-code logical qubits. Then, however, parallel execution of $N$ logical gates requires at least $N$ surface-code logical qubits, leading to large overheads for large $N$. Thus parallel execution of logical gates for qLDPC codes is still challenging.

\section{Summary and outlook}
We have proposed concatenated high-rate quantum error-detecting codes as a new family of high-rate quantum codes for fault-tolerant quantum computation. Since the simple code structure allows for a geometrical interpretation using hypercubes corresponding to logical qubits, we call them many-hypercube codes. The encoding rates are remarkably high, 30\% and 20\% (64 and 256 logical qubits are encoded into 216 and 1296 physical qubits, respectively) for the code distances of 8 and 16, respectively. We have developed a high-performance decoder and fault-tolerant zero-state encoders dedicated to the codes. Using them, we have achieved high error thresholds: 5.6\% for bit-flip errors and 0.9\% for a logical CNOT gate in a circuit-level noise model. Further improvements of the decoder and encoders for the many-hypercube codes are challenging but desirable. We have also explained how to implement logical gates necessary for universal quantum computation. More efficient logical-gate implementations for the many-hypercube codes may be possible but are left for future work. Minimizing the number of logical gates and computational depth for given quantum circuits is also an important compilation problem for the many-hypercube codes.

\begin{acknowledgments}
Parts of numerical simulations were done on the HOKUSAI supercomputer at RIKEN (Project ID Q23614, RB230022).
\end{acknowledgments}

\begin{appendix}

\section{Elementary gates}
\label{gate}
In the $Z$ basis $\{ |0\rangle,|1\rangle \}$, the matrix representations of elementary gates for universal quantum computation are as follows~\cite{3}:

\begin{align}
&\mbox{Identity gate}: 
I=
\begin{pmatrix}
1 & 0 \\ 0 & 1
\end{pmatrix}, 
\nonumber \\
&\mbox{Paul gates}: 
X=
\begin{pmatrix}
0 & 1 \\ 1 & 0
\end{pmatrix}, 
Z=
\begin{pmatrix}
1 & 0 \\0 & -1
\end{pmatrix},
\nonumber \\
& \qquad \qquad \qquad
Y=iXZ=
\begin{pmatrix}
0 & -i \\ i & 0
\end{pmatrix}
\nonumber \\
&\mbox{Hadamard gate}: 
H=
\frac{1}{\sqrt{2}} 
\begin{pmatrix}
1 & 1 \\ 1 & -1
\end{pmatrix}
\nonumber \\
&\mbox{Phase gate}: 
S=
\begin{pmatrix}
1 & 0 \\ 0 & i
\end{pmatrix}
\nonumber \\
&\mbox{Non-Clifford gate}: 
R_Y (\pi/4)=
\begin{pmatrix}
\cos (\pi/8) & -\sin (\pi/8) \\ \sin (\pi/8) & \cos (\pi/8)
\end{pmatrix}
\nonumber \\
&\mbox{CNOT gate}: 
U_{\mathrm{CNOT}}=
\begin{pmatrix}
1 & 0 & 0 & 0 \\ 0 & 1 & 0 & 0 \\ 0 & 0 & 0 & 1 \\ 0 & 0 & 1 & 0
\end{pmatrix}
\nonumber
\end{align}

\section{Hard-decision decoding of the many-hypercube codes}
\label{hard}
The hard-decision decoding of, e.g., the level-3 many-hypercube code is done as follows~\cite{65}. The measurement outcomes of physical qubits in each level-1 block and the corresponding level-1 outcomes are denoted by $\{ x_{i,j,k} | i=1,\ldots,6 \}$ and $\{ x_{i',j,k}^{(1)} | i'=1,\ldots,4\}$, respectively. All the four $x_{i',j,k}^{(1)}$ are set to $F$ (flag indicating an error) if the parity of the six bits, $x_{i,j,k}$, is odd according to the $Z$-stabilizer in Eqs.~(1)--(4). Otherwise, they are set according to the definition of the logical $Z$ operators as follows: $x_{1,j,k}^{(1)}=x_{1,j,k}+x_{2,j,k}$, $x_{2,j,k}^{(1)}=x_{2,j,k}+x_{3,j,k}$, $x_{3,j,k}^{(1)}=x_{4,j,k}+x_{5,j,k}$, and $x_{4,j,k}^{(1)}=x_{5,j,k}+x_{6,j,k}$ (mod 2). Next, using the level-1 outcomes in each level-2 block, $\{ x_{i',j,k}^{(1)} | j=1,\ldots,6\}$, we obtain the corresponding level-2 outcomes, $\{ x_{i',j',k}^{(2)} | j'=1,\ldots,4\}$, as follows. If there is a single $F$ in the six $x_{i',j,k}^{(1)}$, e.g., $x_{i',1,k}^{(1)}$, we can correct this as $x_{i',1,k}^{(1)}=x_{i',2,k}^{(1)}+x_{i',3,k}^{(1)}+x_{i',4,k}^{(1)}+x_{i',5,k}^{(1)}+x_{i',6,k}^{(1)}$ (mod 2) according to the $Z$-stabilizer. (Note that error-detecting codes can correct a located error.) Then, the four $x_{i',j',k}^{(2)}$ are set according to the definition of the logical $Z$ operator, as above. If there is no $F$ and the parity of the six bits is even, we set the four $x_{i',j',k}^{(2)}$ similarly. Otherwise, we set all the four $x_{i',j',k}^{(2)}$ to $F$. Applying this decoding recursively, we finally obtain the logical-level outcomes. If it is $F$, then we randomly choose 0 or 1. For error detection used in the proposed encoders, the decoder returns $F$ (indicating detected errors) unless no $F$ is obtained throughout the decoding.

\section{Symbol-MAP decoding of the many-hypercube codes}
\label{soft}
In the symbol-MAP decoding, we calculate marginal probabilities for each logical qubit, which can be done efficiently. For example, the symbol-MAP decoding of the level-3 many-hypercube code is as follows. Using the prior physical-qubit error probability $p_e$ and the physical-qubit measurement outcomes $\{ x_{i,j,k} | i=1,\ldots,6\}$, the a posteriori probability for a physical-qubit value $x$ is expressed as
\begin{align}
p_{i,j,k}(x) = 
p_e^{I(x\neq x_{i,j,k})} (1-p_e )^{I(x=x_{i,j,k})},
\end{align}
where $I(A)$ is the indicator function that returns 1 if $A$ is true and otherwise 0. Then, the marginal a posteriori probability for a level-1 qubit, e.g., $\mathrm{Q}_{1,j,k}^{(1)}$, is given by
\begin{align}
&p_{1,j,k}^{(1)}(x) 
= 
\frac{\displaystyle \sum_{x_2=0,1} \sum_{x_3=0,1} \sum_{x_4=0,1} 
R_{j,k}^{(1)} (x,x_2,x_3,x_4 )}
{\displaystyle \sum_{x_1=0,1} \sum_{x_2=0,1} \sum_{x_3=0,1} \sum_{x_4=0,1} R_{j,k}^{(1)} (x_1,x_2,x_3,x_4 )},
\\
&R_{j,k}^{(1)}(x_1,x_2,x_3,x_4)
\nonumber \\
&=
\sum_{x'_1=0,1} \sum_{x'_2=0,1} \sum_{x'_3=0,1} \sum_{x'_4=0,1} \sum_{x'_5=0,1} \sum_{x'_6=0,1} \prod_{i=1}^6 p_{i,j,k} (x'_i) 
\nonumber \\
&\qquad \times I\left( \sum_{i=1}^6 x'_i =0 \right) 
I(x_1=x'_1+x'_2) I(x_2=x'_2+x'_3) 
\nonumber \\
&\qquad \times I(x_3=x'_4+x'_5) I(x_4=x'_5+x'_6),
\end{align}
where the summation in the indicator functions is modulo 2. Repeating such calculations recursively, we finally obtain the marginal a posteriori probabilities for the logical-qubit values $\{ x_{i',j',k'}^{(3)} \}$. If $p_{i',j',k'}^{(3)}(0)>0.5$, 
we set $x_{i',j',k'}^{(3)}$ to 0 and otherwise to 1.

\section{Level-by-level minimum distance decoding of the many-hypercube codes}
\label{MD}
The minimum distance decoding is known as one of the highest-performance decoding methods, where we search for a codeword closest to the measurement outcomes in the sense of the Hamming distance. However, there are an exponentially large number of codewords in general, hence it is usually intractable to find such a minimum-distance codeword. To solve this issue, we keep only minimum-distance codewords and discard larger-distance ones at each level from level 1 to the logical level. Since each level-1 codeword [a bit string corresponding to a level-1 computational-basis state, e.g., 000000 or 111111 for $|0000\rangle_L$, see Eq.~(\ref{GHZ})] consists of only six bits, we can easily select minimum-distance codewords and corresponding encoded bit strings (e.g., 0000 for $|0000\rangle_L$) in each level-1 block. While there is a single minimum-distance codeword (encoded bit string) when the parity of the six bits is even, there are six minimum-distance codewords (encoded bit strings) when the parity is odd. To construct level-2 codewords using the level-1 minimum-distance encoded bit strings, we first choose five level-1 blocks from the six ones of each level-2 block, and then pick one of the minimum-distance encoded bit strings from each of the five level-1 blocks. The encoded bit string of the other level-1 block is automatically determined by the parity-check condition corresponding to the $Z$-stabilizer. In general, the determined encoded bit string is not included in the minimum-distance ones of this block. So we evaluate the distance of this encoded bit string, and then obtain the distance of the level-2 block by summing the distances of the six level-1 blocks. We select the minimum-distance codewords (encoded bit strings) of the level-2 block among all the choices of the five level-1 blocks and their encoded bit strings, and keep only them as minimum-distance candidates. This is our strategy to utilize the minimum-distance candidates while satisfying the parity-check condition. Doing this selection recursively, we finally obtain the minimum-distance candidates at the logical level. If we have multiple candidates at the logical level, we randomly choose one of them. See Appendix~\ref{flowchart} for more details of this decoding method.

For error detection used in the proposed encoders, the decoder returns $F$ unless all the numbers of the candidates at the levels from $L_D$ to $L$ are one, where $L$ is the logical level and $L_D$  ($\le L$) is chosen appropriately. 
The most stringent condition for error detection is given by $L_D=1$. To increase the acceptance probabilities at error-detection gadgets in the level-4 zero-state encoder, we relax the condition by setting $L_D$ as $L_D=2$.

\section{Simulation of the bit-flip error model}
\label{bit-flip simulation}
To evaluate the performance of the decoders, we did the following numerical Monte Carlo simulation. We first prepare error-free logical zero states of the many-hypercube code using the encoder in Fig.~\ref{fig1}A. Then independent physical-qubit bit-flip errors are induced with probability $p_{\mathrm{flip}}$ per physical qubit. Finally, we ideally measure the physical qubits in the $Z$ basis. This quantum circuit includes measurements only at the end. Therefore, we can use the fast sampler of Stim~\cite{67} to obtain the physical-qubit measurement outcomes. Then we decode the outcomes using our homemade python codes implementing the above-mentioned decoding methods. If all the logical-qubit outcomes are 0, the decoding succeeds, and otherwise fails. We evaluated the decoding error probabilities by sufficiently many trials of this simulation.

\section{Simulation of logical CNOT gates}
\label{CNOT simulation}
\begin{figure}
    \includegraphics[width=\columnwidth]{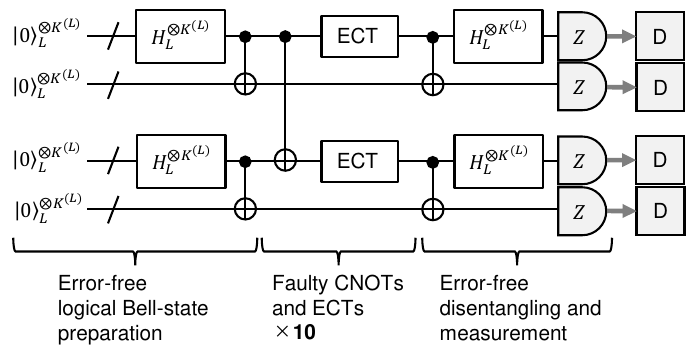}
    \caption{Simulation of logical CNOT gates. We first prepare two sets of error-free logical Bell states encoded with the level-$L$ many-hypercube code in four registers. Then we perform 10 times faulty full transversal logical CNOT gates (implemented by faulty full transversal physical CNOT gates) followed by faulty ECT gadgets on the first and third registers in the circuit-level noise model. Finally, we disentangle the logical states by error-free operations and ideally measure all the physical qubits in the $Z$ basis. The measurement outcomes are decoded by the minimum distance decoder. If all the logical outputs are 0, the 10 sets of full transversal logical CNOT gates succeed, and otherwise fail.}
    \label{fig10}
\end{figure}

To evaluate the performance of full transversal logical CNOT gates in the circuit-level noise model, we did the numerical Monte Carlo simulation shown in Fig.~\ref{fig10}. In this simulation, we first prepare two error-free logical Bell-state blocks using the zero-state encoder in Fig.~\ref{fig1}A and error-free physical CNOT and Hadamard gates. Then, we perform faulty full transversal logical CNOT gates implemented by faulty full transversal physical CNOT gates on the first code blocks of the two Bell-state blocks. This is followed by ECT gadgets in which physical operations are faulty according to the noise model. The faulty logical CNOT gates with faulty ECT gadgets are repeated 10 times. After that, the Bell states are disentangled and converted to the logical zero states by error-free physical CNOT and Hadamard gates. Finally, the logical zero states are ideally measured in the $Z$ basis and the measurement outcomes are decoded by the proposed minimum-distance decoder. Unlike the above simulation of bit-flip errors, this simulation requires mid-circuit measurements and feed-forward operations based on the measurement outcomes. Therefore we use the slow TableauSimulator of Stim~\cite{67} for stabilizer quantum-circuit simulation parts. 
From many trials of this simulation, we estimate the error probability and its standard error of 10 sets of full transversal logical CNOT gates, which are denoted by $p_{10}$ and $\Delta_{10}$, respectively. Then we evaluate those for one set, which are denoted by $p_1$ and $\Delta_1$, as $p_1=1-(1-p_{10} )^{1/10}$ and $\Delta_1=\frac{\Delta_{10}}{10} (1-p_{10} )^{1/10-1}$. Since each set includes $K^{(L)}$ logical CNOT gates at level $L$, we finally evaluate the error probability and its standard error per logical CNOT gate denoted by $p_{\mathrm{CNOT}}$ and $\Delta_{\mathrm{CNOT}}$ as $p_{\mathrm{CNOT}}=1-(1-p_1 )^{1/K^{(L)}}$ and $\Delta_{\mathrm{CNOT}}=\Delta_1/K^{(L)} (1-p_1 )^{1/K^{(L)} -1}$.

\section{Results at level 2 using the Steane method for error detection}
\label{sec-level2}

\begin{figure}
    \includegraphics[width=\columnwidth]{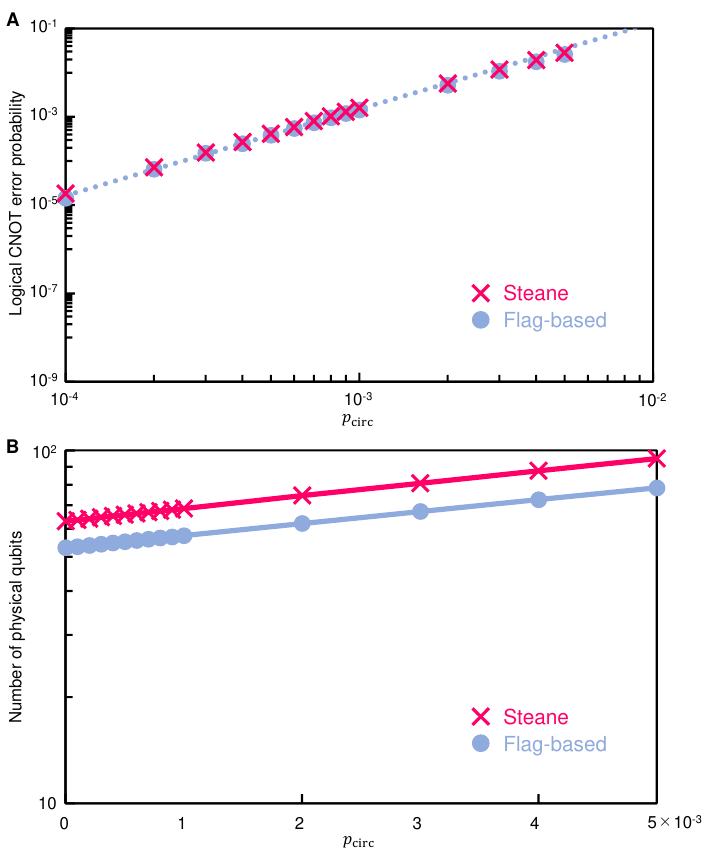}
    \caption{Results at level 4 with the zero-state encoder in Fig.~\ref{fig5}B. (A) Space overheads for the zero-state encoder. (B) Logical CNOT error probability.}
    \label{figS1}
\end{figure}

The crosses in Fig.~\ref{figS1} show the results at level 2 using the Steane method in error-detection gadgets instead of the flag-based method. 
As found from Fig.~\ref{figS1}, the Steane method leads to no performance improvement and a larger space overhead, compared to the flag-based method. This is the reason why we use the flag-based method at level 2.

\section{Results at level 4 using the encoder in Fig. 5B}
\label{encoder5B}

\begin{figure}
    \includegraphics[width=\columnwidth]{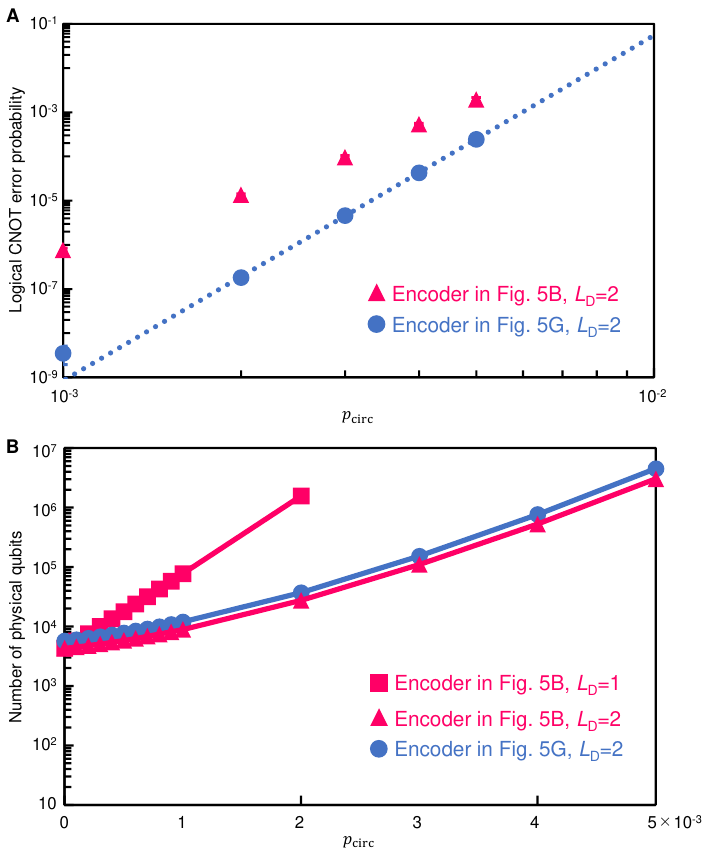}
    \caption{Results at level 4 with the zero-state encoder in Fig.~\ref{fig5}B. (A) Space overheads for the zero-state encoder. (B) Logical CNOT error probability.}
    \label{figS2}
\end{figure}

In this work, we propose the level-4 zero-state encoder in Fig.~\ref{fig5}G to improve the performance compared to the case of the encoder in Fig.~\ref{fig5}B. Here we present the results in the latter case. 
Figure~\ref{figS2}A shows the space overheads at level 4. From this, the encoder in Fig.~\ref{fig5}B with the most stringent condition for error detection, $L_D=1$ (see Appendix~\ref{MD}), leads to the rapid increase of the space overhead with respect to $p_{\mathrm{circ}}$ (shown by squares in Fig.~\ref{figS2}A) due to low acceptance probabilities. Relaxing the condition to $L_D=2$, the overheads (shown by triangles in Fig.~\ref{figS2}A) become similar to those with the encoder in Fig.~\ref{fig5}G (shown in circles in Fig.~\ref{figS2}A and also in Fig.~\ref{fig6}B). However, then the logical CNOT error probability (shown by triangles in Fig.~\ref{figS2}B) is much worse than that with the encoder in Fig.~\ref{fig5}G (shown in circles in Fig.~\ref{figS2}B and also in Fig.~\ref{fig6}A). This is the reason why we propose the encoder in Fig.~\ref{fig5}G at level 4.

\section{Details and flowcharts of the proposed level-by-level minimum distance decoding}
\label{flowchart}
Here we explain our proposed decoding method in more detail using flowcharts. As an example, we focus on the level-3 case.

Figure~\ref{figS3} shows the whole process of the decoding in the level-3 case. Since the subroutines Sub1 and Sub2 are well explained in Appendix~\ref{MD}, here we explain the subroutine Sub3 in detail.

Figure~\ref{figS4} shows the process of Sub3. The subroutine Sub31 for $b_2=6$ is shown in Fig.~\ref{figS5}, where $N_1$ to $N_5$ denote the numbers of the minimum-distance candidates of the first to fifth level-2 blocks, respectively, selected in Sub2. If the total number of candidates, $N_1 N_2 N_3 N_4 N_5$, is larger than a preset threshold $N_{\mathrm{th}3}$, we reduce the number by randomly choosing one of the candidates from the level-2 block with the largest $N$, in order to keep the decoding time short. In this work, we set $N_{\mathrm{th}3}$ to $10^5$. Then we determine the encoded bit string of the sixth level-2 block using the $n_1$-th to $n_5$-th minimum-distance candidates of the first to fifth level-2 blocks, respectively, according to the parity-check condition ($Z$-stabilizer condition). More concretely, the encoded bit string of the sixth level-2 block is determined as follows:
\begin{align}
x_{i',j',6}^{(2)}=x_{i',j',1}^{(2)}+x_{i',j',2}^{(2)}+x_{i',j',3}^{(2)}+x_{i',j',4}^{(2)}+x_{i',j',5}^{(2)},
\end{align}
In general, the level-2 encoded bit string determined above is not included in the minimum-distance candidates of this block selected in Sub2. Therefore, in the subroutine Sub32, we evaluate the distance of this encoded bit string. Figure~\ref{figS6} shows Sub32 for $b_2=6$. In Fig.~\ref{figS6}, $M_1$ to $M_6$ denote the numbers of the minimum-distance candidates of the first to sixth level-1 blocks, respectively, selected in Sub1 in the sixth level-2 block. If $M_1+M_2+M_3+M_4+M_5+M_6$ is larger than a preset threshold $M_{\mathrm{th}2}$, we reduce the number by randomly choosing one of the candidates from the level-1 block with the largest $M$. In this work, we set $M_{\mathrm{th}2}$ to 6. We first determine the encoded bit strings of the five level-1 blocks other than the $b$-th level-1 block in the sixth level-2 block using the $m_b$-th minimum-distance candidate of the $b$-th level-1 block selected in Sub1 and the level-2 encoding bit string determined in Sub31 according to the parity-check condition and the definition of the encoded $Z$ operator. More concretely, when $b=1$, the encoded bit strings of the five level-1 blocks are determined as follows:
\begin{align}
x_{i',2,6}^{(1)}&=x_{i',1,6}^{(1)}+x_{i',1,6}^{(2)},\\
x_{i',3,6}^{(1)}&=x_{i',2,6}^{(1)}+x_{i',2,6}^{(2)},\\
x_{i',5,6}^{(1)}&=x_{i',2,6}^{(1)}+x_{i',1,6}^{(2)}
+x_{i',2,6}^{(2)}+x_{i',3,6}^{(2)}+x_{i',4,6}^{(2)},\\
x_{i',4,6}^{(1)}&=x_{i',5,6}^{(1)}+x_{i',3,6}^{(2)},\\
x_{i',6,6}^{(1)}&=x_{i',5,6}^{(1)}+x_{i',4,6}^{(2)},
\end{align}
In general, the level-1 encoded bit strings determined above are not included in the minimum-distance candidates of the level-1 blocks selected in Sub1. Therefore, we evaluate the distances of the encoded bit strings. This distance evaluation can easily be achieved at level 1. By summing the distances of the six level-1 blocks, we obtain the distance of the sixth level-2 block. In Sub32, we finally select the minimum distance of the sixth level-2 block among all the choices of a level-1 block and its minimum-distance candidates. 
As shown in Fig.~\ref{figS5}, then we evaluate the distances of the level-3 codewords by summing the distances of their six level-2 blocks. We finally select level-3 encoded bit strings with minimum distance, as shown in Fig.~\ref{figS4}.

The decoding of the level-4 many-hypercube code is done in a similar manner to the above level-3 case. In the level-4 case, we set the level-4 threshold $N_{\mathrm{th}4}$ corresponding to the above level-3 threshold $N_{\mathrm{th}3}$ to the same value $10^5$ and the level-3 threshold $M_{\mathrm{th}3}$ corresponding to the above level-2 threshold $M_{\mathrm{th}2}$ to 12.

\end{appendix}

\begin{widetext}

    \begin{figure}[b]
        \includegraphics[width=0.65\columnwidth]{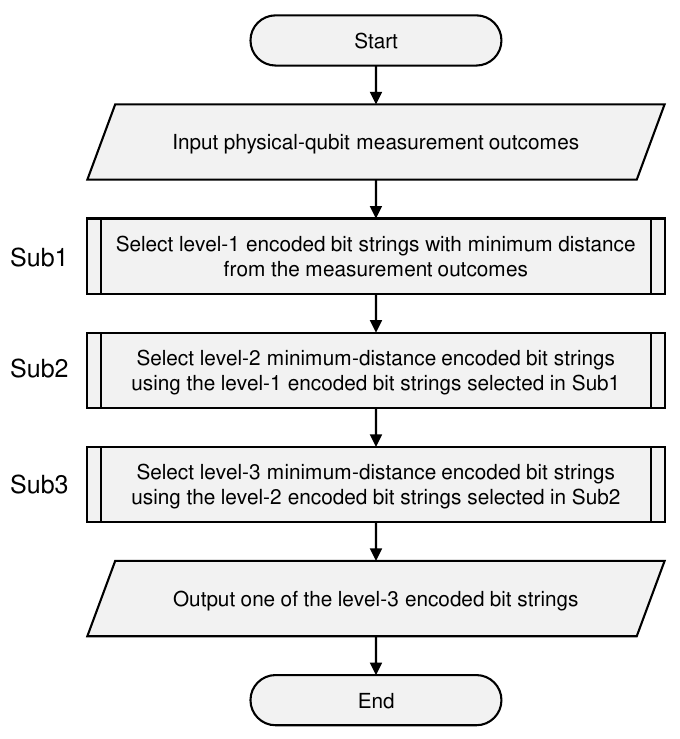}
        \caption{Flowchart of the whole process of the level-by-level minimum distance decoding in the level-3 case. See Fig.~\ref{figS4} for the details of the subroutine Sub3.}
        \label{figS3}
    \end{figure}
    
    \begin{figure}
        \includegraphics[width=0.7\columnwidth]{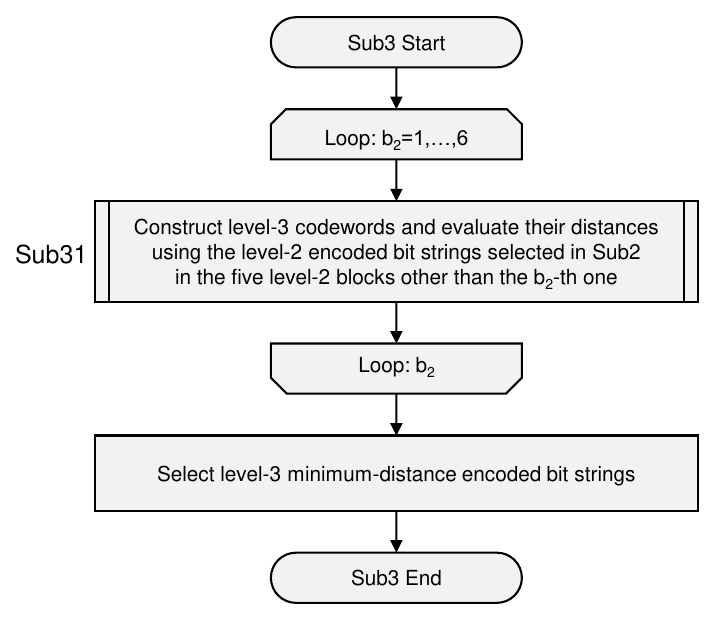}
        \caption{Flowchart of the subroutine Sub3 in Fig.~\ref{figS3}. See Fig.~\ref{figS5} for the subroutine Sub31 when $b_2=6$.}
        \label{figS4}
    \end{figure}

    \begin{figure}
        \includegraphics[width=0.95\columnwidth]{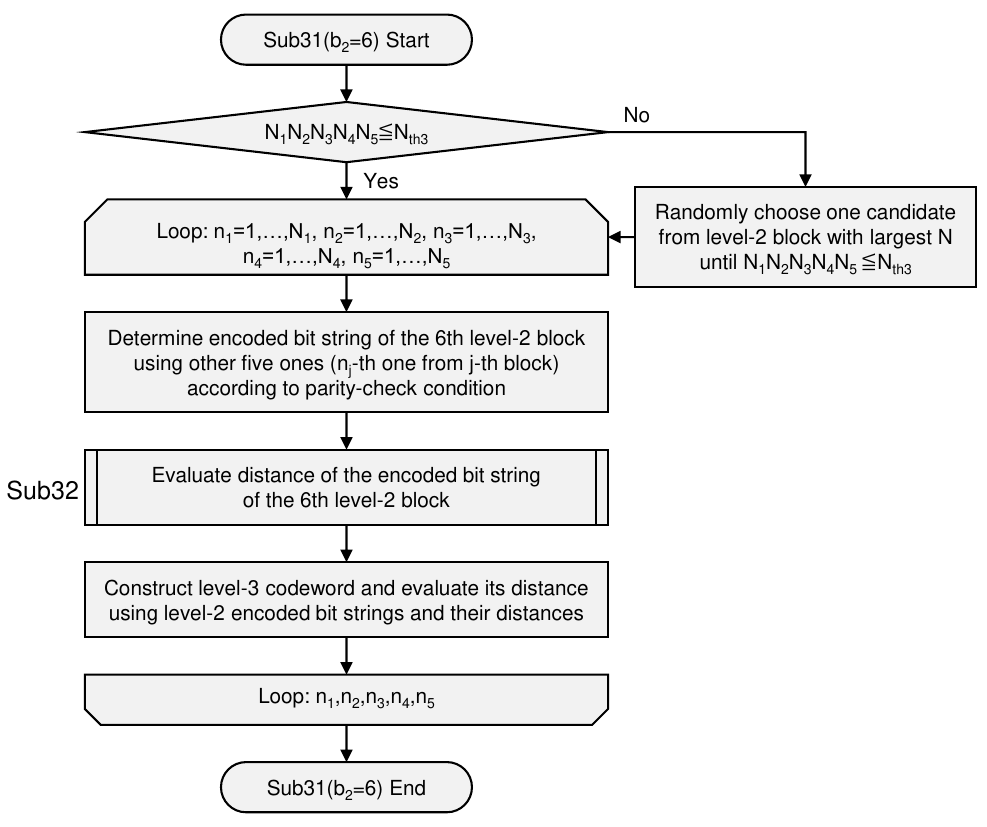}
        \caption{Flowchart of the subroutine Sub31 in Fig.~\ref{figS4} when $b_2=6$. $N_b$ denotes the number of candidates of the $b$-th level-2 block selected in Sub2. See Fig.~\ref{figS6} for the subroutine Sub32.}
        \label{figS5}
    \end{figure}

    \begin{figure}
        \includegraphics[width=0.87\columnwidth]{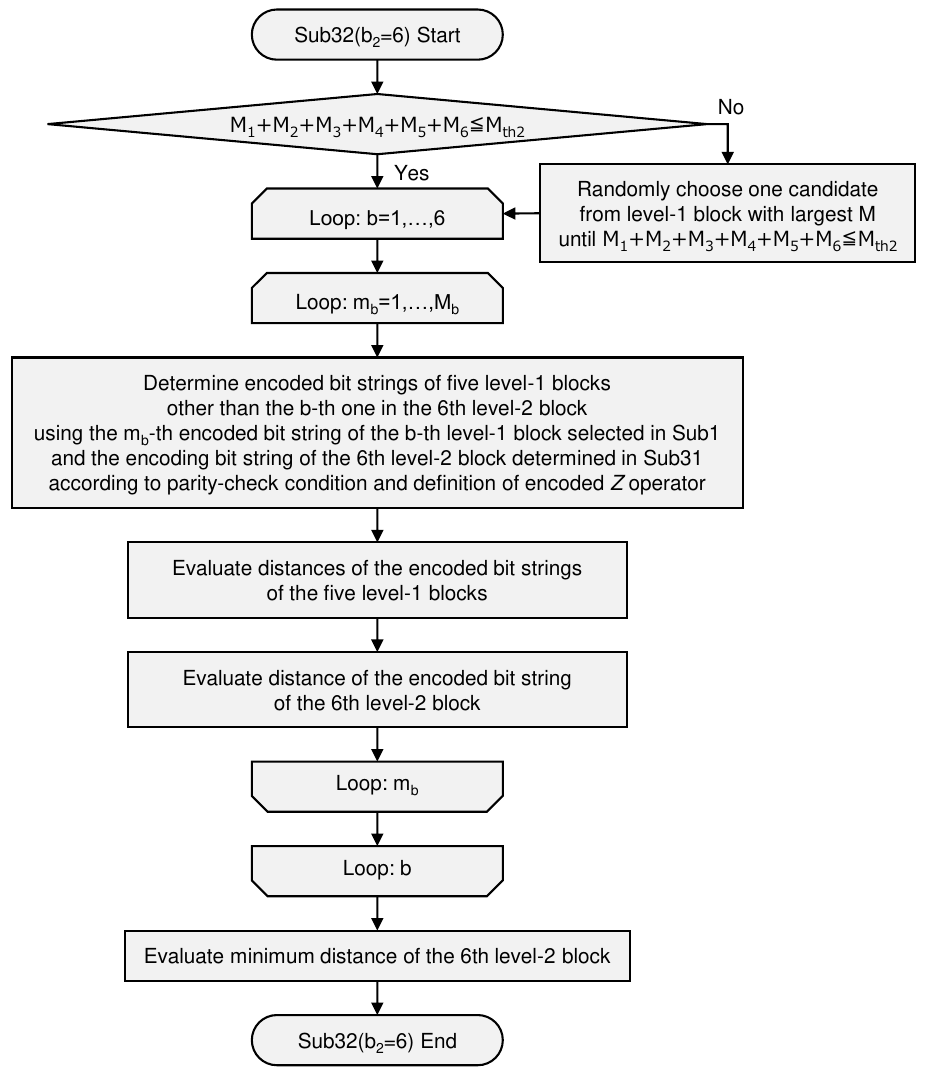}
        \caption{Flowchart of the subroutine Sub32 in Fig.~\ref{figS5}. $M_b$ denotes the number of candidates of the $b$-th level-1 block selected in Sub1.}
        \label{figS6}
    \end{figure}

\end{widetext}

\end{document}